\documentclass[twocolumn,showpacs,preprintnumbers,amsmath,amssymb]{revtex4}

\newcommand{\myskip}[1]{}
\newcommand{\K}{{\cal K}}

\renewcommand{\d}{{\rm d}}

\newcommand{\p}{\partial}

\newcommand{\BEQ}{\begin{eqnarray}}
\newcommand{\EEQ}{\end{eqnarray}}
\newcommand{\BEA}{\begin{eqnarray}}
\newcommand{\EEA}{\end{eqnarray}}
\newcommand{\nn}{\nonumber }
\renewcommand{\d}{{\rm d}}

\newcommand{\mn}{{\mu\nu}}

\usepackage{graphicx}
\usepackage{dcolumn}
\usepackage{bm}

\begin{document}

\title{Exact Schwarzschild-de Sitter black holes in a family of massive gravity models}

\author{ Th. M. Nieuwenhuizen$^{1,2}$}

\email{t.m.nieuwenhuizen@uva.nl}

\homepage{http://staff.science.uva.nl/~nieuwenh/}
\affiliation{
$^{1}$ Center for Cosmology and Particle Physics, New York University, 
 4 Washington Place, New York, NY 10003, USA \\
$^{2}$Institute for Theoretical Physics, University of Amsterdam,
Science Park 904, P.O. Box 94485, Amsterdam, the Netherlands}


\begin{abstract} 

The Schwarzschild-de Sitter and Reissner-Nordstr\"om-de Sitter black hole metrics appear as exact solutions in the
recently formulated massive gravity of  de Rham, Gabadadze and Tolley (dRGT),
where the mass term sets the curvature scale. They occur within a  two-parameter family 
of  dGRT mass terms. They show no trace of a cloud of scalar graviton modes, and in the limit of vanishing graviton mass 
they go smoothly to the Schwarzschild and Reissner-Nordstr\"om metrics.
\end{abstract}

\pacs{ 04.70.Bw,04.20.Cv, 04.20.Jb}

\keywords{Black hole, interior solution}
\maketitle

Ever since Einstein's introduction of General Relativity (GR) in 1915, there has remained the question whether
a mass term for the graviton field can be introduced.  One motivation is theoretical curiosity, but if the solution 
exists, it may also shed a light on the dark energy problem.

 In GR only two physical degrees of freedom are present, the two polarizations of the graviton field.
If a mass term is present, there are also two gravitational vector modes
as for the photon, and two scalar modes. One of the latter is a ``ghost'' field, it has negative
kinetic energy, which would make any solution unstable.  Fierz and Pauli showed that at the quadratic level
a special structure is needed to avoid the ghost~\cite{FierzPauli}. 
van Dam and Veltman~\cite{vanDamVeltman} and also  Zakharov~\cite{Zakharov} 
found that the limit of vanishing mass  does not coincide with GR.
Vainshtein demonstrated that this is caused by a cloud of scalar graviton modes that surround the massive body at an 
intermediately large scale that depends on an inverse power of the graviton mass~\cite{Vainshtein}.
When the mass is made smaller, the cloud moves out further, but it does not disappear, the physical reason for the discontinuity.
Boulware and Deser, however, showed that  the ghost reappears beyond the quadratic level~\cite{BoulwareDeser}.

The program to construct massive gravity was nevertheless solved in last decade. 
In order to start from a general covariant theory of massive gravitation, one first introduces St\"uckelberg fields  \cite{Arkani}. 
This allows a class of potential energies depending on the gravitational metric and an internal Minkowski metric.
In order to prevent the reappearance of the ghost in massive gravity, the set of allowable mass terms is restricted and presented
perturbatively by~\cite{dRG,dRGHP2010}.
The program was advanced recently by de Rham, Gabadadze and Tolley (dRGT)~\cite{dRGT},
who summed up these terms, and found three possible nonlinear combinations:
a quadratic, a cubic and a quartic mass term.
They also show that in the thus obtained theory the Boulware-Deser ghost is absent up to and including fourth order, 
rescuing the approach from its apparent collapse~\cite{Creminelli2005,Mukhanov2010}.

We shall start from dRGT and show that under a certain condition between the coefficients 
of the three mass terms, the Scharzschild black hole survives as a Schwarzschild-de Sitter black hole.
In this solution the decoupling regime remains hidden and there is a continuous transition to the Schwarzschild metric when $m\to0$.

{\it Theory.} 
 The St\"uckelberg fields read in the unitary gauge
$\phi^a=x^a=(ct,\,r\sin\theta\cos\phi,\, r\sin\theta\sin\phi,\,r\cos\theta$).
The Minkowski metric $\eta_{ab}={\rm diag}(1,-1,-1,-1)$  gets represented in spherical coordinates, 

\BEQ \gamma_\mn\equiv \eta_{ab}\p_\mu\phi^a\p_\nu\phi^b,\EEQ
which relates to the Minkowskian background metric
\BEQ \label{Minmet}
\d\sigma^2=\gamma_\mn \d x^\mu\d x^\nu=c^2\d t^2-\d r^2 -r^2\d\Omega^2, 
\EEQ
with $\d\Omega^2=\d\theta^2+\sin^2\theta\d\phi^2$. It represents space in absence of gravitational fields.
Depending on one's view, this is an ``internal space'', an ``auxiliary space'' or a ``pre-existing space",  a space that existed
before gravitational fields grew in it.

When gravitational fields  are present, the static, spherically symmetric metric $\d s^2=g_\mn\d x^\mu\d x^\nu$ reads

\BEQ\label{uvwmetric}
\d s^2=u^2(r)c^2\d t^2-v^2(r)\d r^2-w^2(r)\d\Omega^2.
\EEQ

 de Rham, Gabadadze and Tolley focus on the tensor
 
\BEQ \K^\mu_\nu\equiv \delta^\mu_\nu-(\sqrt{\gamma^{\cdot}_{\cdot}}\,)^\mu_\nu,\qquad 
\gamma^\mu_{\,\,\,\nu}=g^{\mu\rho}\gamma_{\rho\nu},
\EEQ
where $g^\mn$ is the inverse of $g_\mn$ and  for any positive  tensor $A^\mu_\nu$, 
its square root $(\sqrt{A^{\cdot}_{\cdot}})^\mu_\nu$, or, shortly, $\sqrt{A}\,^\mu_\nu$,
 is the solution of $\sqrt{A}\,^\mu_\rho\sqrt{A}\,^\rho_\nu=A^\mu_\nu$ with positive eigenvalues. 
 
In general, let us denote the traces

\BEQ
\K_n={\rm tr}\,\K^n.
\EEQ
In $d=3+1$ dRGT define the terms

\BEA
K_1&=&\K_1,\nn\\
K_2&=&\K_1^2 -\K_2,\qquad \\  
K_3&=&\K_1^3-3\K_1\K_2+2\K_3, \nn \\
K_4&=&\K_1^4 - 6 \K_1^2 \K_2 + 3 \K_2^2 + 8 \K_1\K_3- 6 \K_4. \nn
\EEA
An equivalent definition is

\BEA\label{Keps}
 K_1&=&\frac{-1}{3!}\varepsilon_{\mu\nu\rho\sigma}
\varepsilon^{\dot\mu\nu\rho\sigma}\K^{\mu}_{\,\dot\mu} \nn\\ 
 K_2&=&\frac{-1}{2!}\varepsilon_{\mu\nu\rho\sigma}
\varepsilon^{\dot\mu\dot\nu\rho\sigma}\K^{\mu}_{\,\dot\mu}\K^{\nu}_{\,\dot\nu} \nn\\ 
 K_3&=&\frac{-1}{1!}\varepsilon_{\mu\nu\rho\sigma}\varepsilon^{\dot\mu\dot\nu\dot\rho\sigma}
\K^{\mu}_{\,\dot\mu}\K^{\nu}_{\,\dot\nu} \K^{\rho}_{\,\dot\rho} ,\\
 K_4&=&\frac{-1}{0!}\varepsilon_{\mu\nu\rho\sigma}\varepsilon^{\dot\mu\dot\nu\dot\rho\dot\sigma}
 \K^{\mu}_{\,\dot\mu}\K^{\nu}_{\,\dot\nu} \K^{\rho}_{\,\dot\rho}\K^{\sigma}_{\,\dot\sigma},\nn
 \EEA
where the fully antisymmetric tensor $\varepsilon^{\mu\nu\rho\sigma}$ has
 the element $\varepsilon^{0123}=1/\sqrt{-g}$, where $g=$ det$(g_\mn)$,
 and lowering of its indices is performed with $g_\mn$, so that 
$K_0\equiv (-1/4!)\varepsilon_{\mu\nu\rho\sigma}\varepsilon^{\mu\nu\rho\sigma}=1$.
It is seen that  the $K$'s  are multilinear polynomials.
If $\K^\mu_{\,\,\nu}={\rm diag}(k_0,k_1,k_2,k_3)$ is diagonal, one has

\BEA
\frac{1}{1!}K_1&=&k_0+k_1+k_2+k_3,
\nn\\
\frac{1}{2!}K_2&=&k_0k_1+k_0k_2+k_0k_3+k_1k_2+k_1k_3+k_2k_3,
\nn\\
\frac{1}{3!}K_3&=&k_0k_1k_2+k_0k_1k_3+k_0k_2k_3+k_1k_2k_3,
\nn\\
\frac{1}{4!}K_4&=&k_0k_1k_2k_3. 
\EEA
When the $k$'s represent the eigenvalues of $\K$, these formula's are general.
After all, $K_4/4!$ is the determinant of $\K$, and the other ones can be obtained by differentiating
det$(\epsilon+\K)$ with respect to $\epsilon$ at $\epsilon=0$.

In four dimensional space time no more of such terms exist, while products and powers of the $K$'s 
are not permissible, as they would reintroduce the ghost~\cite{dRGT}. This brings us in the pleasant situation
of a new theory with only a few parameters.
The most general dRGT mass term of the Lagrangian contains these three terms, 

\BEQ\label{Lg=}
L_g=-\frac{m^2c^4}{8\pi G}
\left(\frac{1}{2} K_2+\frac{c_3}{6}K_3+\frac{c_4}{24}K_4\right)-\frac{\Lambda_qc^4}{8\pi G},
\EEQ
where $m=m_gc/\hbar$ is an inverse length scale with $m_g$ being the graviton mass, $c_{3,4}$ 
are dimensionless couplings and $\Lambda_q$ is an explicit cosmological constant, its dimension is 1/m$^2$.
The term $K_1$ can only occur when it is coupled to some source; we shall not consider such a situation.
The Lagrangian (\ref{Lg=}) is added to  the Hilbert-Einstein action 

\BEQ \label{Ltot=}
L_{\rm tot}=-\frac{R}{16\pi G}+L_g, \EEQ
and possible other Lagrangians from other matter fields, such as electromagnetism.
The full Lagrangian (\ref{Ltot=}) is invariant under combined coordinate transformations
and gauge transformations of the St\"uckelberg field~\cite{Arkani,Berezhiani}.

{\it Energy conservation.}  
The energy momentum tensor 
 $T^\mn_g={(-2}/{\sqrt{-g}}){\delta (\sqrt{-g}L_g)}/{\delta g_\mn}$ reads for diagonal $\K$
 
 \BEA 
T^\mn_g =-g^\mn[L_g +(1-k_\mu)\frac{\p L_g}{\p k_\mu}].
 \EEA
 In the case (\ref{Minmet}), (\ref{uvwmetric}) we shall have

\BEA k_0 = 1-\frac{1}{u},\quad k_1=1-\frac{1}{v},\quad k_2=k_3=1-\frac{r}{w}.\EEA
The key idea of this Letter is to restrict ourselves to solutions that obey
 
 \BEQ
w(r)=a_1 r,\qquad k_2=k_3=1-\frac{1}{a_1}.
\EEQ
The energy conservation $(T_g)^\mu_{\,\,\nu;\mu}=0$ imposes for $\nu=r$

\BEA 
P[u,u',v,r]\left[c_3 - \frac{1 - 2 a_1}{1-a_1}\right]=c_4 -\frac{ 1 - 3 a_1 + 3 a_1^2}{(1 - a_1)^2},
\EEA
for a rational $P[u,u',v,r]$. We may write the solution as

\BEQ a_1=\frac{b_1}{1+b_1},\qquad c_3=b_1-1,\qquad c_4=1-b_1+b_1^2. 
\EEQ
A special role of this case was noticed in an application to cosmology~\cite{dRGHP2010,note}.
Now for any $u(r)$ and $v(r)$ the energy momentum tensor acts as a cosmological constant,

\BEQ (T_g)^\mu_\nu=\Lambda_c\delta^\mu_\nu,\quad
\Lambda_c\equiv\Lambda_q+\Lambda_g,\quad
\Lambda_g\equiv-\frac{m^2}{b_1}.\EEQ
Stability of the solution imposes that $\Lambda_g<0$~\cite{dRGHP2010}, so that  $b_1>0$ is required
in the case of our interest, $m^2>0$.

In fact, the argument continues to apply when all considered parameters are functions of time.
Energy conservation will then impose that only $\Lambda_c$ be a constant.
This may open the road to models where $\Lambda_q$ denotes the naive cosmological constant from quantum physics,
which gets largely canceled by a negative term from massive gravity, not due to fine tuning but due to energy conservation.
This setup could then leave a small $\Lambda_c$ as net result. 

{\it de Sitter space}. 
The de Sitter metric

\BEA u&=&\sqrt{1-\frac{1}{3}\Lambda_cw^2},\quad v=\frac{a_1}{\sqrt{1-\frac{1}{3}\Lambda_cw^2}},
\EEQ
where $w=a_1r$, solves the Einstein equations

\BEQ G^\mn\equiv R^\mn-\frac{1}{2}Rg^\mn=\frac{8\pi G}{c^4}T^\mn_g,
\EEQ
 because in terms of the variable $w$ the only effect of $L_g$
is the cosmological constant $\Lambda_c$. (The normalization of $u$ is fixed to $u\to1$ for $\Lambda_c\to0$).
There are three cases: $\Lambda_c=0$ because the external and the induced cosmological constants cancel each other, 
this occurs for  $a_1=a_1^c\equiv m^2/(\Lambda_q+m^2)$, where space time is just Minkowskian;
the anti-de Sitter universe $\Lambda_c<0$ ($a_1<a_1^c$) where the solution makes sense globally and, finally,
the de Sitter universe $\Lambda_c>0$ ($a_1>a_1^c$) which has a singularity at $w_{\rm dS}=a_1r_{\rm dS}\equiv\sqrt{3/\Lambda_c}$.
In our massive gravity this is a physical singularity.

For the same reasons the black hole with geometrized mass  $M$ 
and charge $Q$, extended with a de Sitter term,

\BEQ \label{MQsol}
u=\sqrt{1-\frac{2M}{w}+\frac{Q^2}{w^2}-\frac{1}{3}\Lambda_cw^2},\qquad v=\frac{a_1}{u},\EEQ
is an exact solution of the Einstein-Maxwell equations.
The case $Q=0$ is called the Schwarzschild-de Sitter metric,
its physical mass is $Mc^2/G$ (so that the Schwarzschild radius equals  $2M$ when $\Lambda_c=0$). 
The general case is called the Reissner-Nordstr\"om-de Sitter solution.
It has a charge $q$, vector potential $A_\mu=(q/w,0,0,0)$ and in SI units an electromagnetic 
energy density $q^2/8\pi\epsilon_0w^4\equiv (Q^2/2w^4)(c^4/8\pi G)$, where $e^2/4\pi\epsilon_0\hbar c=1/137$ if $q=e$,
is the proton charge. $Q=qc^{-2}\sqrt{2G/\epsilon_0}$ is the geometrized charge, it has dimension of length.

The solution  (\ref{MQsol}) has an energy density~\cite{LandauLifshitz,BabakGrishchuk,NEPLRTG2007,NEPLBH2008}

 \BEA t^{00}\!=\!\frac{-2 a_1^6c^4 ( \frac{ Q^2}{w^2} - \frac{ M}{ w}+\frac{1}{3}\Lambda_c w^2  )
 (\frac{ Q^2}{w^2} - \frac{ 2M}{ w}-\frac{1}{3}\Lambda_c w^2 ) }
 {8\pi Gw^2(1-\frac{ 2M}{ w} +\frac{ Q^2}{w^2} -\frac{1}{3}\Lambda_c w^2)^2},\quad
  \EEA
and further

\BEQ t^{11} = \frac{a_1^4c^4}{8\pi Gw^2} ( \frac{ Q^2}{w^2}+\Lambda_c w^2 )
( \frac{ Q^2}{w^2}- \frac{ 2M}{ w}-\frac{1}{3}\Lambda_c w^2  ),
\EEQ
while the other $t^\mn$ vanish.  The total energy momentum tensor involves  $\gamma={\rm det}(\gamma_\mn)$ 
and reads~\cite{BabakGrishchuk,NEPLRTG2007}

\BEQ
\Theta^\mn=t^\mn+\frac{\Lambda_cc^4g}{8\pi G\gamma}\,g^\mn=t^\mn+\frac{\Lambda_cc^4a_1^6}{8\pi G}\,g^\mn.
\EEQ
The total energy density $\Theta^{00}$ goes quadratically to $-\infty$ near the Schwarzschild horizon $M+\sqrt{M^2-Q^2}+{\cal O}(\Lambda_c)$, 
as noticed before~\cite{NEPLBH2008}, signaling the well known peculiarities of the horizon; it goes quadratically to $+\infty$ near the 
de Sitter horizon $w_c=\sqrt{3/\Lambda_c}$, which will require a long time to establish in practice.

Since $\Theta^{00}$ is a physical quantity in our massive gravity theory, not a gauge dependent quantity like in GR,
one may wonder whether such infinities should be permissible in a physical theory, or are indeed allowed as an
infinite time limit. An attempt to construct black hole-type solutions without horizon
was reported in ref.~\cite{NEPLBH2008}.

The above approach can be applied to Carter's generalization of the Kerr-Newman metric to de Sitter space~\cite{Carter1973},

\BEA
\d s^2&=&
\frac{\Delta}{\Sigma {\cal E}_0^2}\left[\d t-a\sin^2\theta\d\phi\right]^2
-\frac{\Sigma\,\d w^2}{\Delta}
-\frac{\Sigma\,\d\theta^2}{{\cal E}_\theta }
\nn\\
&&-\frac{{\cal E}_\theta \sin^2\theta}{\Sigma {\cal E}_0^2}\left[a\d t-(w^2+a^2)\d\phi\right]^2.
\EEA
with $ {\cal E}_\theta=1+\frac{1}{3}\Lambda_c a^2\cos^2\theta$, ${\cal E}_0=1+\frac{1}{3}\Lambda_c a^2$ and

\BEA
\Sigma&=& w^2 + a^2 \cos^2\theta,\quad 
\\
\Delta& = & (w^2 + a^2)\left[1-\frac{1}{3}\Lambda_cw^2\right] -2Mw+Q^2.  
 \nn
\EEA
One has to determine,  analytically or numerically, the eigenvalues of $(g^{\mu\rho}+\delta g^{\mu\rho})\gamma_{\rho\nu}$ and then of $\K^\mu_\nu$, 
for general or small $a$ or $\Lambda_c$.  But
$T_g^\mn$ appears to deviate from $\Lambda_c\,g^\mn$ whenever $a\neq0$. This non sequitur is puzzling.

{\it Summary and outlook}.
We have presented exact, inhomogeneous solutions that 
show no trace of a cloud of scalar graviton modes;  in the limit of vanishing graviton mass 
they go smoothly to the Schwarzschild and Reissner-Nordstr\"om metrics.
Apparently, both the decaying and growing Yukawa solutions of the decoupling limit occur, and they conspire
to give at the non-linear level the discussed smooth metrics.
This finding is in agreement with the decoupling of the scalar mode on a
cosmological  de Sitter background~\cite{dRGHP2010}.

The solution has infinities in the physical energy density at the Schwarzschild and de Sitter horizons.
Whether they are physically permissible may be elucidated by a study of the fluctuations.

The existence of exact black hole solutions, that are continuous in the limit where the
graviton mass vanishes, gives a special role for the considered two-parameter family of massive gravity models.
One may study whether in these models also solutions exist that go to the Minkowski 
metric at distances beyond $1/m$. Other interesting questions for future concern the description of stars and time-dependence.
Finally, it is interesting to know whether our restricted class of massive gravity theories is physically more relevant than the general case.

The considered massive gravity is very appealing. Its Einstein equations involve only second order time-derivatives so the Cauchy problem is well 
posed: the evolution is determined by the usual initial conditions. The theory involves a few parameters, $G$, $m$, $b_1$ and $\Lambda_q$.
It is a gauge theory that allows no other terms, like $R^2$ or $K_2K_3$, since  this would either spoil the Cauchy problem or reintroduce the ghost.
In particle physics related arguments led to the 't Hooft -- Veltman renormalizability of quantum gauge theories  and then to the standard model.
If one tries to quantize the present theory with Newton's constant having the ``wrong'' dimension, a non-perturbative renormalization scheme
seems the most to strive for.

\acknowledgments
It is a pleasure to thank Gregory Gabadadze for introducing me to the topic, for discussions and reading the manuscript in draft.
I also thank Giga Chkareuli for discussion.

\end{document}